\documentstyle{article}

\newcommand {\tsub}[1]{_{\mbox{\protect\scriptsize #1}}}

\newcommand {\Ref}[1]{(\ref{#1})}
\newcommand
{\Stretch}[1]{\renewcommand{\baselinestretch}{#1}\large\normalsize}
\newcommand {\figureStretch}{\Stretch{0.9}}

\newcommand {\unStretch}{\Stretch{1.2}}

\newcommand {\bm}[1]{{\mbox{\boldmath $#1$}}}
\newcommand {\leqsim}{\widetilde{<}}
\newcommand {\geqsim}{\widetilde{>}}

\begin{document}

\pagestyle{empty}

\Stretch{3}
\begin{center}

{\large 24/1/95}
\vspace*{2em}

{\huge \bf Density matrix renormalisation group study of the correlation
function of the bilinear-biquadratic spin-1 chain}
\vspace*{5em}

\Stretch{1.2}
{\large
R.\ J.\ Bursill$^{\dagger}$, T.\ Xiang$^{*}$ and
G.\ A.\ Gehring$^{\dagger}$
}
\vspace*{2em}

$\dagger$
Department of Physics,
The University of Sheffield,
S3 7RH, Sheffield, United Kingdom.\\
Email: r.bursill@sheffield.ac.uk

$*$ Interdisciplinary Research Center in Supeconductivity, The
University of
Cambridge CB3 0HE, United Kingdom.\\
\end{center}

\vfill
\eject

\Stretch{1.2}

\begin{abstract}

Using the recently developed density matrix renormalization group
approach, we study the correlation function of the spin-1
chain with quadratic and biquadratic interactions. This allows us to
define and calculate the periodicity of the ground state which differs
markedly from that in the classical analogue. Combining our results with
other studies, we predict three phases in the region where the quadratic
and biquadratic terms are both positive.

\end{abstract}

\setcounter{page}{0}
\vfill
\eject

\pagestyle{plain}

\section{Introduction}
\label{Introduction}
\setcounter{equation}{0}

The application of the real space renormalisation group (RSRG) to
quantum lattice models \cite{Wilson} has received renewed interest in
recent times due to the development of the density matrix
renormalisation group (DMRG) by White and coworkers \cite{development}.
DMRG studies of the spin-1 chain with pure antiferromagnetic exchange
\cite{White}, \cite{Sorenson} yielded ground state properties with
unprecedented reliability and accuracy. Applications to coupled spin
\cite{coupled_spin_chains} and Hubbard \cite{coupled_hubbard_chains}
chains have followed and applications to fermion systems in two
dimensions are being persued \cite{fermion_systems}.

The spin-1 chain has been the subject of much recent interest because of
theoretical predictions \cite{Haldane} and experimental observations
\cite{experimental} of a gap in the exitation spectrum of integer spin
Heisenberg models, verified beyond doubt in the spin-1 case in
\cite{White}.

In this paper we present results from the application of the DMRG to a
generalisation of the standard spin-1 chain where a biquadratic
interaction is added to the usual Heisenberg exchange viz
\begin{equation}
{\cal H}=\sum_{i}
\left[
\cos\gamma S_{i}.S_{i+1}
+
\sin\gamma\left(S_{i}.S_{i+1}\right)^{2}
\right]
\label{H}
\end{equation}
where $\bm{S}_{i}$ is the spin-1 operator for site $i$ and $\gamma$ is a
parameter determining the relative strength of the bilinear and
biquadratic interactions. We concentrate on the regime
$0\leq\gamma\leq\pi/2$ where the quadratic term is antiferromagnetic and
the biquadratic term opposes any kind of order between neighbours.

As mentioned, in the $\gamma=0$ case the existence of a gap $\Delta$
(between the spin-0 ground state and the spin-1 first exited state) has
been well established. Also calculated in \cite{White} is the spin-spin
correlation function
\begin{equation}
C(i-j)=
\left<
S_{i}^{z}S_{j}^{z}
\right>
\label{correlation_function}
\end{equation}
which is found to decay exponentialy with alternating sign, being well
fitted by
\begin{equation}
C(r)\sim
A\frac{(-1)^{r}e^{-r/\xi}}
{r^{\frac{1}{2}}}
\label{Haldane_asymptotic}
\end{equation}
with correlation length $\xi\approx 6.03$.
The ground state is disordered in the conventional sense although an
interesting type of topological long range order, characterised by a
{\em string order
parameter}
\begin{equation}
g(l)
\equiv
\left<
S_{0}^{z}
\left(
\prod_{j=1}^{l-1}e^{i\pi S_{j}^{z}}
\right)
S_{l}^{z}
\right>
\end{equation}
exists and has been calculated \cite{White}.

For $\gamma=\arctan(1/3)$ the ground state wavefunction can
be written down \cite{Affleck}, having a simple valence-bond solid
structure. The ground state energy is known exactly and a gap to the
first exited $(S=1)$ state can be shown to exist. The correlation
function is given by
\begin{equation}
C(r)=\frac{4}{3}(-1)^{r}3^{-r}\mbox{ for }r>0
\label{Affleck_correlation}
\end{equation}
so that $C(r)$ again decays exponentially with alternating sign and has
correlation length
\begin{equation}
\xi=1/\log 3=0.9102\ldots
\label{Affleck_correlation1}
\end{equation}

At $\gamma=\pi/4$ the model is exactly solvable by Bethe ansatz
\cite{Lai_Sutherland}, the exitation spectrum being gapless and having
soft modes at 0 and $\pm 2\pi/3$. In this case standard field theory
predicts that $C(r)$ should decay algebraically. This suggests that a
phase transition occurs at
$\gamma=\gamma\tsub{c}\in\left(\arctan(1/3),\pi/4\right]$ where the
correlation length diverges.

Two of us (T.\ X.\ and G.\ A.\ G.) have studied the ground
state energy density $\epsilon\tsub{G}$ of \Ref{H} using a variant of
the RSRG \cite{Xiang}. It was noticed qualitatively that there was a
crossover from conventional antiferromagnetic period-2 behaviour to
period-3
behaviour in the (weak) chain length dependence of $\epsilon\tsub{G}$ as
$\gamma$ was varied between 0 (the Heisenberg point) and $\pi/4$ (where,
as a result of the soft mode at $2\pi/3$, a triply periodic ground state
in expected). It was found that the crossover occurred near
$\gamma=0.15\pi$ though we emphasise that this observation was only
qualitative.

The following questions thus arise naturally:
\begin{enumerate}
\item
Is $\gamma\tsub{c}=\pi/4$?
\item
What are the exponents which characterise the divergence of the
correlation length and the subsequent algebraic decay of $C(r)$?
\item
Does $\Delta$ vanish for $\gamma\geq\gamma\tsub{c}$?
\item
What is a good definition of the {\em periodicity} of the ground state?
\item
How does the periodicity vary with $\gamma$?
\item
Is there a relationship between the periodicity and the classical pitch
angle $\theta^{*}$
(the angle between sucessive spins in the ground state of the classical
($S=\infty$) analogue)?
\item
How should the string order parameter be defined (for general $\gamma$)
and what is its value?
\end{enumerate}

The above questions have been the subject of both analytical
\cite{Affleck1}--\cite{Nomura} and numerical
\cite{Fath_Solyom}--\cite{Reed} studies, often with conflicting results.

Though there are no completely firm answers to any of the above
questions, the most definitive study of \Ref{H} is the most recent study
by F\'{a}th and S\'{o}lyom \cite{Fath_Solyom2} where finite size scaling
and twisted boundary condition techniques are applied in analysing exact
results from small chains. The answer to question 1 is found to be in
the affirmative, there being a transition of the
Kosterlitz-Thouless type at $\gamma=\gamma\tsub{c}=\pi/4$. It is also
found, in answer to question 3, that $\Delta$ does indeed vanish for
$\gamma\tsub{c}\leq\gamma\leq\pi/2$. The exponent $\sigma$,
characterising the opening of the gap is also estimated.

As mentioned, the questions concerning the periodicity of the ground
state were qualitatively addressed in \cite{Xiang}. They were also
studied qualitatively in \cite{Fath_Solyom} where again evidence of a
triply periodic ground state was found for $\pi/4\leq\gamma\leq\pi/2$
and also in a region below $\pi/4$. This agrees qualitatively with
analyical predictions from spin wave \cite{Xian} and variational
\cite{Nomura} theories.

In this paper we attempt to answer 4-6 quantitatively by calculating
\Ref{correlation_function} using the DMRG. The DMRG allows us to study
much longer chains than those studied in
\cite{Fath_Solyom}--\cite{Fath_Solyom2}. This facilitates the accurate
calculation of $C(r)$ for $r$ up to around $50$ and hence its Fourier
transform
\begin{equation}
\tilde{C}(q)\equiv\sum_{r}C(r)e^{iqr}
\label{Fourier}
\end{equation}
We are then able to define and calculate the periodicity of the ground
state in terms of the position where $\tilde{C}(q)$ has its peak.

In the next section we describe our implementation of the DMRG for
\Ref{H} and our results for $C(r)$ and the periodicity of the ground
state.

\section{DMRG results for the spin-spin correlation function}

\setcounter{equation}{0}

\subsection{The DMRG method for spin chains}

As mentioned, the DMRG was introduced in a series of papers
\cite{development} where efficient algorithms for calculating low-lying
energies and correlation functions of spin chains are
described in great detail. We will therefore be very brief in our
description of the method. We restrict our discussion to the infinite
lattice algorithm \cite{development} which we used for our calculations.

The DMRG is an iterative, truncated basis procedure whereby a large
chain (or superblock) is built up from a single site by adding a small
number of sites at a time. At each stage the superblock consists of
system and enviroment blocks (determined from previous iterations) in
addition to a small number of extra sites. Also determined from previous
iterations are the matrix elements of various operators such as the
block Hamiltonians and the spin operators for the sites at the end(s) of
the blocks) with respect to a truncated basis. Tensor products of the
states of the system block, the enviroment block and the extra sites are
then formed to provide a truncated basis for the superblock. The ground
state $\left|\psi\right>$ (or other targeted state) of the superblock is
determined by a
sparse matrix diagonalization algorithm.

At this point, correlation functions, local energies and other
expectation values are calculated with respect to $\left|\psi\right>$.
Next, a
basis for an augmented block, consisting of the system block and a
specified choice of the extra sites, is formed from tensor products of
system block and site states. The augmented block becomes the system
block in the next iteration. However, in order to keep the size of the
superblock basis from growing, the basis for the augmented block is
truncated. We form a density matrix by projecting
$\left|\psi\right>\left<\psi\right|$ onto
the augmented block which we diagonalise with a dense matrix
routine. We retain the {\em most probable} eigenstates (those
with the largest eigenvalues) of the density matrix in order to form a
truncated
basis for the augmented block that is around the same size as the
system block basis. Matrix elements for the Hamiltonian and active site
operators, together with any other operators that are required for say,
correlation functions are then updated.

The environment block used for the next iteration is usually chosen to
be a reflected version of the augmented block. The initial system and
environment blocks are chosen to be single sites.

We note that the key parameter determining the accuracy and,
correspondingly, the computer requirements (both cpu time and memory) is
$n\tsub{s}$, the number of states retained per block (of good quantum
numbers) at each iteration. $n\tsub{s}$ therefore determines the
truncation error, which is the sum of the eigenvalues of the density
matrix corresponding to states which are shed in the truncation process.
A large truncation error indicates that too many important states are
being shed and that $n\tsub{s}$ is too small. The error in quantities
such as the ground state energy scale linearly with the truncation error
\cite{White}.

The application of the DMRG to the spin-1 chain with pure
antiferromagnetic nearest neighbour exchange ($\gamma=0$) is described
in detail in \cite{White}. We will not repeat this description here,
rather we describe the results of our extension of the method to
$\gamma\neq 0$, paying attention to particular difficulties that arise
in this case.

\subsection{Application of the DMRG to the bilinear-biquadratic chain}

We have applied the infinite lattice DMRG algorithm to the calculation
of $C(r)$ for \Ref{H} using a number of superblock configurations and
boundary conditions. All the interactions (intrablock, interblock and
superblock Hamiltonians) commute with the total $z$ spin
$S^{z}\tsub{T}\equiv\sum_{i}S^{z}_{i}$, so $S^{z}\tsub{T}$ is a good
quantum number which can be used to block diagonalize the system,
enviroment and super blocks. The ground state of the superblock
$\left|\psi\right>$
is a singlet with zero total spin so we only need to consider superblock
states with $S^{z}\tsub{T}=0$.

With the computing power available to us (14 megaflop Silicon Graphics
Indigo II workstation with 80 megabytes of free memory) we found that
the most cpu efficient configuration was an open ended superblock of the
form system-site-enviroment. We gauged the efficiency of a given
configuration by assessing how rapidly the ground state energy density,
a by-product of the calculation, converged to its known value in cases
where exact values or high precision numerical results are available.

With $n\tsub{s}=40$, a calculation for the $\gamma=0$ case involving 49
iterations (99-site superblock) takes around 14 hours of cpu time and
requires around 5 megabytes of memory. With $n\tsub{s}=50$ for
$\gamma=\pi/4$, 250 iterations took up 10 days of cpu time. In general
the resource requirements increase as the square of $n\tsub{s}$. We next
describe the attainable accuracy which, as we shall see, depends
strongly on $\gamma$ through the exitation gap and rate of decay of the
correlation functions.

\subsection{Testing of the method}

\subsubsection*{Ground state energy density}

We calculate $\epsilon\tsub{G}$ by evaluating the local energy
\cite{White}
\begin{equation}
\cos\gamma S_{i}.S_{i+1}
+
\sin\gamma\left(S_{i}.S_{i+1}\right)^{2}
\end{equation}
where $i$ is a site in the middle of the superblock (ie the end site of
the system block). This reduces the end effects considerably. For a
given choice of $n\tsub{s}$ the scheme is iterated until this quantity
converges. The local energy is calculated for various values of
$n\tsub{s}$ and the error is estimated using the proportionality of the
error to the truncation error \cite{White}.

\begin{description}
\item{Haldane point $(\gamma=0)$}

In this case the ground state energy density is known to high precision
(at least seven significant figures) from the large scale DMRG
calculation using careful smoothing of the boundary conditions
\cite{White} which has been verified independently by a very large scale
exact diagonalization and finite size scaling analysis \cite{Golinelli}.
In our implementation, agreement to seven significant figures can be
reached using $n\tsub{s}=40$ and around $50$ iterations are required for
the local energy to converge to this accuracy.

\item{Affleck point $\tan\gamma=1/3$}

As mentioned, the ground state wavefunction has a simple, known
structure and the exact ground state energy density is known. With
$n\tsub{s}=40$ we recover the ground state energy density to within
round off error after only a few iterations. This is due to the fact
that the ground state can be built up exactly by retaining only a few
states at each iteration.

\item{Lai-Sutherland point $\gamma=\pi/4$}

As mentioned, the energy spectrum in this case is gapless and the
correlations are predicted to decay algebraically. These two conditions
are very unfavourable for accurate DMRG calculations \cite{development}
(convergence is algebraic rather than exponential in $n\tsub{s}$ if the
spectrum is gapless) and this is borne out in our implementation. That
is,
with $n\tsub{s}=50$ we achieve agreement with the exact ground state
energy density only to within one tenth of one percent.
The Lai-Sutherland point is critical so boundary effects are strong and
around 250 iterations are required to achive convergence of the local
energy to four significant figures.


\end{description}

Table~\ref{energies} shows the DMRG ground state energy density together
with the exact or precise result for the three cases mentioned above. In
each instance the exact or precise result was recovered within the error
bound calculated from the truncation error.




\subsubsection*{Finite chains}

We also tested the algorithm by calculating the exact ground state
energy
and correlation functions for small chains (of up to 13 sites) by the
Lanzcos method and ensuring that the DMRG recovered the exact result.

\subsection{Results for the correlation function}

To calculate the correlation functions we retain and update the matrix
elements of the $S^{z}_{i}$ for $i$ near the active end of the blocks at
each iteration. This part of the calculation that uses the bulk of the
cpu and memory resources. We calculate $C(r)$ by evaluating
$\left<\psi\right|S_{i}^{z}S_{j}^{z}\left|\psi\right>$ for $j-i=r$ where
$i$ and $j$ are sites
equidistant from the centre of the superblock \cite{White} (again to
minimise end effects and optimize convergence).

There are two sources of error in calculating
$\tilde{C}(q)$---truncation of the Hilbert space and truncation of the
Fourier series.  In forming the Fourier transform \Ref{Fourier} we use
the Fourier coefficients $C(r)$ which converge to at least two
significant figures. The convergence rate of $C(r)$ with lattice size
decreases with $r$ because the number of truncations performed on the
spin matrices used to form $C(r)$ increases linearly with $r$. The first
few coefficients have similar dependence on $n\tsub{s}$ (or the
truncation error) to the ground state energy density. For
$r\geqsim 10$ errors due to spin matrix truncation, rather than
truncation of $\left|\psi\right>$, begin to become dominant.

Plots of $\tilde{C}(q)$ versus $q$ for various values of $\gamma$ using
$n\tsub{s}=40$ are given in Fig.~\ref{C_tilde(q)}. At the Haldane point
$\tilde{C}(q)$ is peaked sharply around $q=\pi$ as is to be expected
from strongly period-2 behaviour. As $\gamma$ is increased, the width of
the peak increases. At some point $\gamma=\tilde{\gamma}$ the peak
begins to
move away from $\pi$ towards $2\pi/3$ as $\gamma$ is increased towards
$\pi/4$, the Lai-Sutherland point, where period-3 behaviour is expected.
The peak sharpens as $\gamma$ approaches $\pi/4$ consistent with the
algebraic decay predicted in $C(r)$ at $\gamma=\pi/4$.


\subsubsection*{Estimates of the correlation length}

In Fig.~\ref{correlation_length} we give a crude estimate of the
correlation length $\xi$ obtained simply by linearly fitting
$\log |C(r)|$ over around twenty sites. This procedure is meaningful if
$\tilde{C}(q)$ has a strong peak at $\pi$. The procedure can also be
made more explicit if $\tilde{C}(q)$ has a strong peak at or near
$2\pi/3$.
For example, for $\tan\gamma=3/4$, $\tilde{C}(q)$ has a peak at
$0.6718(4)\pi$. $C(r)$ is plotted as a function of $r$ in
Fig.~\ref{C(r)_tang_=_3/4} where we see strong triply periodic
behaviour. Plots of $\log|C(3r)|$, $\log|C(3r+1)|$ and $\log|C(3r+2)|$
are given in Fig.~\ref{logC(r)s}. The slopes of the three curves are
roughly equal and the fitting procedure yields the result of taking the
average of these slopes.




$\xi$ estimated in this way depends weakly on range of sites over which
the average is taken though in general significant systematic error is
expected due to Hilbert space truncation in addition to error in not
taking into account the (generally unknown) correct asymptotic form of
$C(r)$. At the Haldane point (with $n\tsub{s}=40$) we get
$\xi\approx 5.2$---only 87 percent of its precise value $6.0\ldots$
\cite{White} which was obtained from a very large scale calculation,
assuming and taking into account the extra algebraic factor in the
asymptotic form \Ref{Haldane_asymptotic}. At the Affleck point we
recover the exact correlation length to five significant figures though
this is again due to the simplicity of the wavefunction.

For $0\leq\tan\gamma\leqsim 3/4$ $C(r)$ decays exponentially
and the error due to truncation of the Fourier series is negligible. For
$3/4\leqsim\tan\gamma\leq 1$ the correlation length is large and
there is significant oscillation in $\tilde{C}(q)$ due to Fourier series
truncation. This can be seen in Fig.~\ref{C_tilde(q)} in the
$\tan\gamma=0.8$ case. For $\gamma\geq\pi/4$, $C(r)$ appears to decay
algebraically and the peak appears to remain precisely at $2\pi/3$. This
is consistent with previous numerical studies \cite{Fath_Solyom} where
gapless excitations with soft modes at $\gamma=0,\pm2\pi/3$ are found
and it is conjectured that the model is critical and triply periodic in
the whole region. Our crude estimates of $\xi$ in the region
$0\leq\tan\gamma\leq 0.8$ are consistent with the prediction of
\cite{Fath_Solyom2} that $\gamma\tsub{c}=\pi/4$ and hence that $\xi$
should diverge as $\gamma\rightarrow\pi/4^{-}$.

\subsubsection*{Peak position and periodicity of the ground state}

{}From the above discussion we are lead to the natural (working)
definition of the ground state periodicity, namely $2\pi/q^{*}$ where
$q^{*}$ denotes the position where $\tilde{C}(q)$ has its peak. The peak
position converges quite rapidly with $n\tsub{s}$ and also with
$n\tsub{F}$, the number of converged Fourier coefficients retained, in
cases where Fourier series truncation is not negligible. In
table~\ref{peak_convergence} we list $q^{*}$ for various values of
$n\tsub{F}$ and $n\tsub{s}$ for the case where $\tan\gamma=3/4$.




In Fig.~\ref{periodicity} we plot $q^{*}$ as a function of $\gamma$. As
mentioned above, the ground state seems to be triply periodic in the
whole region $\pi/4\leq\gamma<\pi/2$. The fluctuations of $q^{*}$ about
$2\pi/3$ being around half of one per cent which is within the error
that can be estimated from the rate of convergence. This is in agreement
with predictions from spin wave \cite{Xian} and variational
\cite{Nomura} theories.


We see that the periodicity seems to be very close to 3 in the region
$0.225\pi\leqsim\gamma\leq\pi/2$. It is not clear from our data whether
$q^{*}$ lies strictly above $2\pi/3$ in the whole open interval
$0<\gamma<\pi/4$ or not. Significantly higher precision in $q^{*}$ would
be required to decide this.

We next consider the point $\gamma=\tilde{\gamma}\approx 0.13\pi$,
beyond which the
periodicity is greater than 2. In the vicinity of $\tilde{\gamma}$ the
correlation length is very short and convergence of $q^{*}$ with
$n\tsub{s}$ is very rapid. The error (between our best result and
results for smaller $n\tsub{s}$) is roughly proportional to the
truncation error. As with the ground state energy density, we use the
truncation error to extrapolate to $n\tsub{s}=\infty$ and give an
estimate of the error which is around $0.01$ percent. At
$\tan\gamma=0.438$ we have $q^{*}=\pi$ and at $\tan\gamma=0.4381$ we
have $q^{*}/\pi=0.9951(1)$ so $\tilde{\gamma}$ lies between these
points.
$\tilde{C}''(\pi)$, the second moment at $q=\pi$ is very well fitted by
a linear function in the region $0.4<\tan\gamma<0.5$. Finding the point
where the least squares fit vanishes then yields
$\tilde{\gamma}=0.43806(4)$.

\subsubsection*{Comparison with the classical model}

In the classical analogue of \Ref{H}, the quantum spin operators $S_{i}$
become classical spin variables which can take on any value on the unit
sphere. The energy is
\begin{equation}
E=\sum_{i}\epsilon\left(\theta_{i}\right)
\end{equation}
where
$\epsilon(\theta)=\cos\gamma\cos\theta+\sin\gamma\cos^{2}\theta$ and
$\theta_{i}$ is the angle between $S_{i}$ and $S_{i+1}$. This is
minimized for
\begin{eqnarray}
\theta_{i}
& \equiv &
\theta^{*}
\\
& = &
\left\{
\begin{array}{ll}
\pi & 0\leq\gamma\leq\tilde{\gamma}\tsub{cl}
\\
\arccos\left(-\left(\cot\gamma\right)/2\right)
&
\tilde{\gamma}\tsub{cl}\leq\gamma\leq\pi/2
\end{array}
\right.
\end{eqnarray}
where $\tilde{\gamma}\tsub{cl}\equiv\tan^{-1}\frac{1}{2}$ denotes the
position of the onset of the classical spiral phase and $\theta^{*}$
denotes the classical pitch angle.

In the N\'{e}el phase
$0\leq\gamma\leq\tilde{\gamma}\tsub{cl}$ the classical
ground state has the trivial degeneracy of fixing the direction of one
of the spins. In the spiral phase
$\tilde{\gamma}\tsub{cl}<\gamma<\pi/2$ the
classical ground state has further degeneracy in that the azimuthal
angle between successive spins can take on any value. The classical
correlation function is obtained by averaging $S_{i}.S_{j}$ over all
classical ground states \cite{Flory} viz
\begin{equation}
C(r)=\frac{1}{3}\left(\cos\theta^{*}\right)^{r}
\end{equation}

Classically then, $\tilde{C}(q)$ has a quadratic peak at $q=\pi$ in the
whole region $0\leq\gamma\leq\pi/2$. If we, however, restrict the spiral
states
so that the $S_{i}$ lie in a plane and make angle $i\theta^{*}$ with
$S_{0}$ then we have $C(r)=\cos\left(r\theta^{*}\right)$ and
$\tilde{C}(q)$ has delta function peaks at $q=\theta^{*}$ and
$2\pi-\theta^{*}$, vanishing elsewhere.

We include a plot of $\theta^{*}$ as versus $\gamma$ in
Fig.~\ref{periodicity}. Clearly, quantum fluctuations break the
degeneracy in the classical spiral state forcing successive spins to lie
in a plane. Moreover, the $\theta^{*}$ and $q^{*}$ curves have marked
differences---$\tilde{\gamma}\neq\tilde{\gamma}\tsub{cl}$ and $q^{*}$
remains
fixed at $2\pi/3$ for all $\gamma>\pi/4$. The classical solution is
therefore highly unstable against incommensurate distortions of spins.
Indeed \Ref{H} is one of the simplest examples of a quantum model where
the $1/S$ expansion away from the classical solution fails
\cite{Gehring}.

\section{Summary and conclusions}

We have applied the DMRG to calculate the ground state correlation
function and to define and calculate the periodicity of the ground state
of the bilinear-biquadratic spin-1 chain \Ref{H} in the frustrated
regime where both coupling constants are positive.

Combining our study with those of F\'{a}th and S\'{o}lyom, we predict
that in the region studied (where the quadratic term is
antiferromagnetic and the biquadratic term opposes alignment of
neighbouring spins) the model has three phases:
\begin{enumerate}
\item
$0\leq\gamma\leq\tilde{\gamma}$

In this phase the model has short ranged
antiferromagnetic order, $\tilde{C}(q)$ having a quadratic maximum at
$q=\pi$ and the ground state having periodicity 2.
\item
$\tilde{\gamma}<\gamma<\gamma\tsub{c}$

In this phase the model has short ranged spiral order, $\tilde{C}(q)$
having a quadratic peak at $q=q^{*}<\pi$. We define the periodicity of
the ground state to be $2\pi/q^{*}$.
\item
$\gamma\tsub{c}\leq\gamma<\pi/2$

In this phase the model has quasi-long range order, $C(r)$ decaying
algebraically. The ground state has period 3, $\tilde{C}(q)$ having a
cusp singularity at $q=2\pi/3$.
\end{enumerate}

We estimate $\tan\tilde{\gamma}=0.43806(4)$. The predicition of
\cite{Fath_Solyom2} that $\gamma\tsub{c}=\pi/4$ is not inconsistent with
our results for the peak position $q^{*}$ and our crude estimates of the
correlation length.

The authors are grateful for J.\ Parkinson and Y.\ Xian from UMIST and
E.\ Shender from Oxford for useful discussions and also to M.\ E.\ Cates
from Cambridge for informing us of reference \cite{Flory}. R.\ J.\ B.\
would like
to acknowledge the support of EPSRC grant no. GR/J26748.

\vfill
\eject

\section*{Table Captions}

\noindent 1. Estimates of the ground state energy density
$\epsilon\tsub{G}$ from the DMRG and exact or precise results.

\noindent 2. Convergence of the peak position $q^{*}$
for the case where $\tan\gamma=3/4$.

\vfill
\eject

\setcounter{table}{0}

\figureStretch
\begin{table}[htbp]
\centering

\begin{tabular}{||c|c|c||}
\hline
$\tan\gamma$ & DMRG & Exact \\
\hline
0 & $-1.4014845(38)$ & $-1.40148403\ldots$ \\
$1/3$ & $0.666666666667(1)$ & $2/3$ \\
1 & $0.2963(8)$ & $0.29678\ldots$ \\
\hline
\end{tabular}

\caption{}
\label{energies}
\end{table}
\unStretch

\figureStretch
\begin{table}[htbp]
\centering

\begin{tabular}{||c|c|c||}
\hline
$n\tsub{s}$ & $n\tsub{F}$ & $q^{*}$ \\
\hline
20 & 20 & 0.67610 \\
20 & 30 & 0.67362 \\
20 & 40 & 0.67314 \\
20 & 50 & 0.67285 \\
30 & 20 & 0.67552 \\
30 & 30 & 0.67301 \\
30 & 40 & 0.67238 \\
30 & 50 & 0.67195 \\
40 & 20 & 0.67526 \\
40 & 30 & 0.67275 \\
40 & 40 & 0.67221 \\
40 & 50 & 0.67182 \\
\hline
\end{tabular}

\label{peak_convergence}
\caption{}
\end{table}
\unStretch

\vfill
\eject

\section*{Figure Captions}







\setcounter{figure}{0}

\figureStretch
\begin{figure}[htbp]
\caption{
$\tilde{C}(q)$ versus $q$ for various values of
$\gamma$.}
\label{C_tilde(q)}
\end{figure}
\unStretch

\figureStretch
\begin{figure}[htbp]
\caption{
Estimate of the correlation length $\xi$ for various
values of $\gamma$.}
\label{correlation_length}
\end{figure}
\unStretch

\figureStretch
\begin{figure}[htbp]
\caption{
The correlation function $C(r)$ in the case where
$\tan\gamma=3/4$.}
\label{C(r)_tang_=_3/4}
\end{figure}
\unStretch

\figureStretch
\begin{figure}[htbp]
\caption{
$\log|C(3r)|$ (solid curve), $\log|C(3r+1)|$ (dashed
curve) and $\log|C(3r+2)|$ (dot-dashed curve) in the case where
$\tan\gamma=3/4$.}
\label{logC(r)s}
\end{figure}
\unStretch

\figureStretch
\begin{figure}[hbtp]
\caption{
$q^{*}$ (dots) and $\theta^{*}$ (solid line) as
functions of $\gamma$.}
\label{periodicity}
\end{figure}
\unStretch

\end{document}